\documentclass{aa}  
\usepackage{graphicx}
\usepackage{txfonts}
\usepackage{lipsum}
\usepackage{subcaption}
\usepackage{lscape}
\usepackage{placeins}
\usepackage[]{hyperref}

\newcommand{\sgra}{\mbox{Sgr A$^\ast$}}
\newcommand{\dg}{\mbox{$^{\circ}$}}
\newcommand{\pcmq}{\mbox{cm$^{-2}$}}
\newcommand{\pcmc}{\mbox{cm$^{-3}$}}
\newcommand{\psec}{\mbox{s$^{-1}$}}
\newcommand{\psr}{\mbox{sr$^{-1}$}}
\newcommand{\pmev}{\mbox{MeV$^{-1}$}}

\newcommand{\pflux}{\mbox{photons~\pcmq~\psec}}
\newcommand{\eflux}{\mbox{erg~\pcmq~\psec}}
\newcommand{\elum}{\mbox{erg~\psec}}
\newcommand{\fmevster}{\mbox{ph~\pcmq~\psec~\pmev~\psr}}
\newcommand{\gray}{\mbox{$\gamma$-ray}}

\begin{document}

\title{Detection of positron in-flight annihilation from the Galaxy}

\author{J.~Kn\"odlseder\inst{1} \and
             K.~Sabri\inst{1,2} \and
             P.~Jean\inst{1} \and
             P.~von Ballmoos\inst{1} \and
             G.~Skinner\inst{3} \and
             W.~Collmar\inst{4}}

\institute{Institut de Recherche en Astrophysique et Plan\'etologie, Universit\'e de Toulouse, CNRS, CNES, UPS,
9 avenue Colonel Roche, 31028 Toulouse, Cedex 4, France\\
              \email{jknodlseder@irap.omp.eu}
              \and Laboratoire Univers et Particules de Montpellier, Universit\'e de Montpellier, CNRS/IN2P3, 
Place Eug\`ene Bataillon, 34095 Montpellier Cedex 5, France
              \and Honorary Senior Research Fellow, School of Physics and Astronomy, University of Birmingham, UK
              \and Max-Planck-Institut f\"ur extraterrestrische Physik, Postfach 1603, 85740 Garching, Germany}

\date{Received 23 June 2025 / Accepted 9 July 2025}

\abstract
{Although the annihilation of positrons towards the Galactic centre was established more than 
50 years ago through the detection of a 511~keV \gray\ line, the origin of the positrons remains 
unknown.
The \gray\ line should be accompanied by continuum emission from positron in-flight annihilation, 
which, until now, has not been detected.}
{We aim to detect positron in-flight annihilation emission, as it provides information on the kinetic 
energy of the positrons that is key in determining the origin of Galactic positrons.} 
{We analysed archival data obtained by the COMPTEL instrument on the Compton Gamma-Ray
Observatory satellite to search for positron in-flight annihilation emission in the MeV energy range.}
{Our analysis revealed extended emission in the MeV energy range towards the bulge of the Galaxy, 
which we attribute to in-flight annihilation of positrons produced with kinetic energies of $\sim2$~MeV.
The observed spectrum suggests that positrons are produced quasi-mono-energetically, which
could occur by the annihilation of dark matter particles with masses of $\sim3$~MeV, or
through bulk motion in the jet of the microquasar \object{1E~1740.7--2942}.
We furthermore detected a point-like emission component in the MeV energy range towards the 
Galactic centre that is the plausible low-energy counterpart of the Fermi-LAT source 
\object{4FGL~J1745.6--2859}.
The broad band spectrum of the source may be explained by the injection of pair plasma from 
the supermassive black hole \object{\sgra} into the interstellar medium, which would also explain the
point-like 511~keV line emission component that was discovered by INTEGRAL/SPI at the
Galactic centre.}
{The observed positron in-flight annihilation spectrum towards the Galactic bulge excludes 
$\beta^+$ decays from radioactive isotopes, as well as any mechanism producing highly relativistic 
positrons as the origin of the Galactic bulge positrons.}


\maketitle
\nolinenumbers

\section{Introduction}

Observations of the Galactic centre region in the 1970s revealed a 511~keV \gray\ line 
and continuum emission below 511~keV, which was attributed to two- and three-photon annihilation of 
thermalised positrons \citep{johnson1972,leventhal1978}.
The most recent observations with the Spectrometer on INTEGRAL (SPI) suggest the existence of
several positron annihilation emission components, including extended emission from the Galactic 
bulge that is slightly offset from the Galactic centre, a point-like emission component
at the Galactic centre with a characteristic scale of no more than a few hundred parsecs,
and extended emission from the Galactic disc with poorly constrained 
morphology \citep{bouchet2010,skinner2014,siegert2016}.

Despite progress on the characterisation of the emission, the origin of the positrons remains an 
open question, with many possible channels for positron production discussed in the 
literature \citep{prantzos2011,churazov2020}.
The initial kinetic energy of positrons varies strongly between these channels, yet observations of the 
511~keV line (or the three-photon continuum emission) probe thermalised positrons that have lost
their initial energy due to interactions with ambient medium, erasing the link to their sources.
In addition, during the slowdown, positrons can travel substantial distances from the production 
site \citep{jean2009}, blurring the link between source distribution and emission morphology.

Positrons may also annihilate in flight before having slowed down, giving rise to Doppler-shifted 
two-photon continuum emission in the energy interval from $m_e c^2 / 2$ to 
$E_p +  m_e c^2 \, 3/2$, where $m_e c^2$ is the rest energy and $E_p$ is the kinetic energy of the 
positrons.
For strongly relativistic positrons, the fraction of in-flight annihilations in the interstellar medium (ISM)
may reach several tens of percent, implying that a noticeable flux from positron in-flight annihilation is
expected \citep{agaronioan1981}.
Previous attempts to place limits on the positron injection energy include an analysis of 
INTEGRAL/SPI data that provided non-constraining upper limits \citep{churazov2011},
and comparisons of expected in-flight \gray\ emission with estimates of observed flux
from within a defined patch of sky that place an upper limit of 3--7.5~MeV
on the injection energy, depending on the ionisation state of the ISM \citep{beacom2006,sizun2006}.

The latter limits were based on published spectra of Galactic diffuse emission obtained using 
observations with the Compton Telescope (COMPTEL), an instrument operating on the Compton 
Gamma-Ray Observatory (CGRO) during 1991--2000.
Since COMPTEL did not have direct imaging capabilities, data analysis relied on forward folding
of sky distributions with the instrument response function, which allowed the reconstruction of images and
spectra in the 0.75--30 MeV energy range \citep{schoenfelder1993}.
Furthermore, the instrumental background, which far exceeds any celestial signal, was
modelled using an iterative approach, introducing some circularity that makes the resulting images
and spectra dependent on the assumed celestial intensity distribution \citep{bloemen1999}.
In particular, emission components that were not explicitly searched for will be suppressed,
which implies that upper limits on the positron injection energy derived from
published COMPTEL spectra of Galactic ridge emission are not very constraining.

To overcome this limitation, we performed the first dedicated analysis of COMPTEL data that
explicitly searched for the signal from positron in-flight annihilation on top of the emission
known to be associated with the Galactic ridge.
This analysis was carried out as part of an effort to reduce the environmental footprint of astrophysical 
research by addressing science questions using existing data \citep{knoedlseder2022b,knoedlseder2024}.
We find that such an analysis enables the identification of continuum emission, one component of 
which has a morphology and spectrum, suggesting that it is due to in-flight annihilation from 
positrons.\footnote{
All analysis results are available in numerical form for download at \url{https://zenodo.org/records/15814023}.
}

\section{Observations and analysis methods}

\subsection{Data preparation}

For our study, we used COMPTEL data that are publicly available at NASA's High Energy Astrophysics 
Science Archive Research Center (HEASARC).
Only data recorded before the second reboost of the CGRO spacecraft were used, because the second 
reboost resulted in a significant increase in the instrumental background at low energies, substantially 
decreasing the instrument sensitivity \citep{weidenspointner1999,ryan2023}.
COMPTEL data are split into so-called viewing periods, and our analysis covered 215 viewing 
periods, starting from viewing period 1.0 and ending with viewing period 616.1, spanning from 
16 May 1991 to 18 March 1997.

COMPTEL observed the sky in the energy range from 0.75~MeV to 30~MeV, and we analysed the data 
using either the full energy range or one of two sub-bands, spanning 0.75--3~MeV and 3--30~MeV, 
respectively.
The sub-bands were split at 3~MeV because around that energy there is a distinct
change in the spatial morphology of the emission from the Galactic bulge.
We split the data in each energy band into logarithmically spaced energy bins, excluding the 
1.7--1.9~MeV energy range that contains the signal from the 1.8~MeV radioactive decay line of $^{26}$Al.
In total, 16 energy bins were used for the full energy band to accurately determine any spectral cut-offs
in the emission, while five and four energy bins were used for the low and high sub-bands, respectively. 
There is one more energy bin for the low sub-band compared to the high sub-band due to 
the exclusion of the 1.7--1.9~MeV interval from the 1.5--2.1~MeV energy bin, which split this energy bin 
into two.

For each energy bin, events were binned per viewing period into standard COMPTEL data cubes 
spanned by Compton scatter direction $(\chi,\psi)$ and Compton scattering angle $\bar{\varphi}$.
Bin sizes of $\Delta\chi=\Delta\psi=1\dg$ and $\Delta\bar{\varphi}=2\dg$ were used throughout the analysis.
For each energy bin, we combined data cubes for individual viewing periods into a single data cube
that was centred on the Galactic centre and spanned $\pm120\dg$ in Galactic longitude and 
$\pm70\dg$ in Galactic latitude.
This large data cube provides a good contrast between the inner Galactic plane region, which is rich in 
Galactic \gray\ photons, and high-latitude regions that are essentially free of such photons,
allowing a clear separation of \gray\ point sources, Galactic ridge emission, and any
excess emission at the Galactic centre.
At the same time, the data cube excludes events from the Crab nebula and pulsar which is by far 
the brightest MeV \gray\ source, avoiding any interference in our analysis by events 
from this source.

\subsection{Data analysis}

All data analysis was performed using version 2.1.0 of the ctools \citep{ctools2016} and GammaLib 
\citep{gammalib2011} software packages.
We analysed the data by adjusting model components to the observed events using a maximum-likelihood 
method.
Each model component consists of parametrised spatial and spectral models, which we fitted
jointly to the data cubes for all energy bins in a given energy band.
We used the {\tt comlixfit} script for model fitting, which derives a model of the instrumental background 
event distribution in the data cubes while simultaneously adjusting the parameters of the celestial 
model components \citep{knoedlseder2022a}.
The most critical parameter of the algorithm is $N_{\rm incl}$, which specifies the number of 
$\bar{\varphi}$ layers of the three-dimensional data cubes used to estimate the background
model.
As demonstrated by \cite{knoedlseder2022a}, large values of $N_{\rm incl}$ conserve the celestial
source fluxes but may lead to substantial residuals attributed to an inadequate modelling of the
instrumental background event distribution.
Smaller values of $N_{\rm incl}$ reduce these residuals, but do so at the expense of suppressing 
some of the source flux and detection significance.
To minimise background residuals while still conserving source fluxes, we used 
$N_{\rm incl}=11$ in this study.
We used the BGDLIXF algorithm for background modelling, which is an evolution of the BGDLIXE
algorithm described in \cite{knoedlseder2022a} and which properly treats background event
rate variations between and within viewing periods (see Appendix \ref{sec:background}).

We modelled the celestial \gray\ emission using components for the in-flight positron annihilation 
emission, ten point sources previously detected in MeV \gray s in the examined
region (see Appendix \ref{sec:sources}), and components for Galactic ridge and cosmic \gray\ emission.
We modelled the spectra of all point sources and the cosmic \gray\ background using power 
laws, while we used more elaborate spectral forms for the other components.
We adjusted the spectral parameters of all components during model fitting, with the exception of the 
cosmic \gray\ background component, which is spatially isotropic and therefore cannot be 
distinguished from the instrumental background through spatial model fitting.
We therefore fixed this component to the spectrum
$I(E) = 1.12 \times 10^{-4} ( E/{5 \, {\rm MeV}} )^{-2.2}$ \fmevster, as
suggested by \cite{weidenspointner1999}.

\begin{figure}[!t]
\centering
\includegraphics[width=\columnwidth]{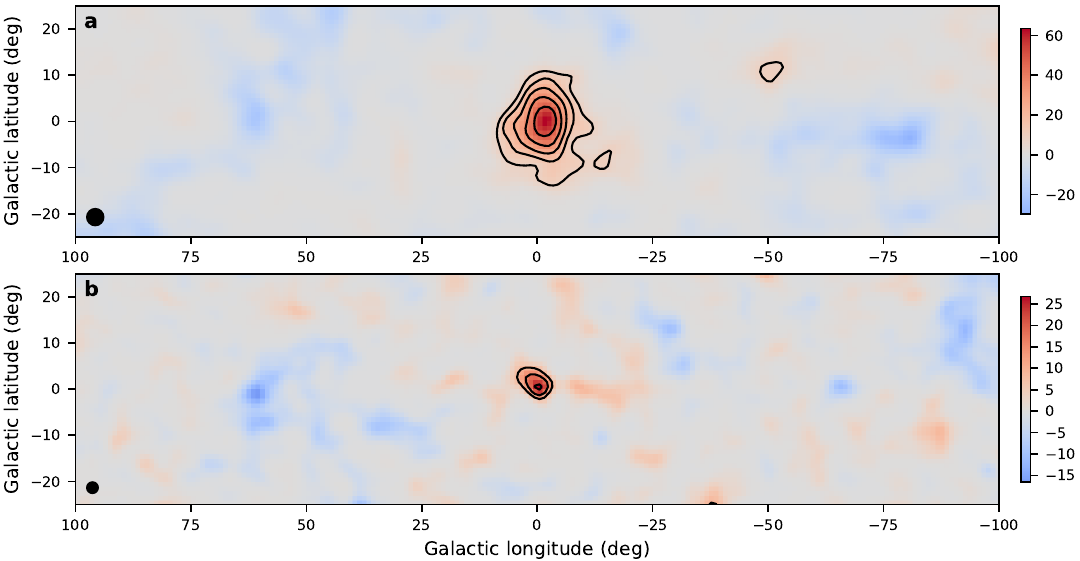}
\caption{
Test statistic (TS) maps in the 0.75--3~MeV (a) and 3--30~MeV (b) energy bands.
The significance, expressed in Gaussian sigma, corresponds to $\sqrt{\rm TS}$, where negative TS 
values correspond to model fits that resulted in negative source flux.
The contours indicate the pre-trial detection significance in Gaussian sigma, with contours spaced 
by $1\sigma$ and the first contour starting at $3\sigma$.
The filled circles in the lower-left corners of the maps indicate the 68\% containment region of the 
COMPTEL point spread function.
\label{fig:tsmaps}
}
\end{figure}

The Galactic ridge emission at energies in the MeV range is believed to originate primarily from 
diffuse inverse-Compton (IC) emission of cosmic-ray electrons, yet contributions from unresolved 
source populations cannot be excluded \citep{bloemen1999,strong2000b}.
To avoid any biases from an assumed spatial distribution of Galactic diffuse emission, 
and to allow for the possible presence of unresolved point sources,
we implemented an empirical model using two-dimensional (2D) asymmetric Gaussian functions 
for the spatial distribution.
For the low- and high-energy band fits, we used a single 2D asymmetric Gaussian function as the 
spatial model, with a power-law spectral model for which we adjusted the prefactor and index.
For the full-energy band fits we used two 2D asymmetric Gaussian functions to
accommodate different spatial morphologies of the Galactic ridge emission at low and high
energies. We modelled the spectra of both components using log-parabola functions, adjusting the 
prefactor, the index, and the curvature during the fit.

\section{Results}

\subsection{Excess emission maps}

We searched for excess emission towards the Galactic bulge region by generating test statistic 
(TS) maps spanning $200\dg\times50\dg$ in Galactic longitude and latitude around the Galactic centre,
following the method described in \cite{knoedlseder2022a}.
We derived the maps by fitting a point-source model to a $1\dg$ grid of source locations, on top of a 
reference model that includes all known \gray\ sources in the region of interest, comprising Galactic 
ridge emission, point sources, and cosmic \gray\ background, as well as a description of the instrumental 
background.
As the morphology of the excess emission changes within the COMPTEL energy range, we derived 
separate maps for the $0.75-3$~MeV and $3-30$~MeV energy bands, which are shown in 
Fig.~\ref{fig:tsmaps}.

The TS map for the $0.75-3$~MeV energy band reveals significant excess emission towards the 
Galactic centre, with a morphology that is clearly extended and an emission peak that appears
slightly offset towards negative longitude.
This is the first report of MeV excess emission from the Galactic bulge, and the 
resemblance of the TS map to the initial 511~keV line emission maps from INTEGRAL/SPI 
\citep{knoedlseder2005} is striking.
The map shows few features outside the Galactic bulge excess, indicating that the reference model 
adequately describes the emission from the Galactic ridge and the known point sources within the 
analysis region.
Significant excess emission is also seen in the $3-30$~MeV energy band, with a point-like morphology 
located at the Galactic centre.
Besides the Galactic centre source, there is no other significant excess in the map.

\begin{figure*}[!t]
\centering
\includegraphics[width=\textwidth]{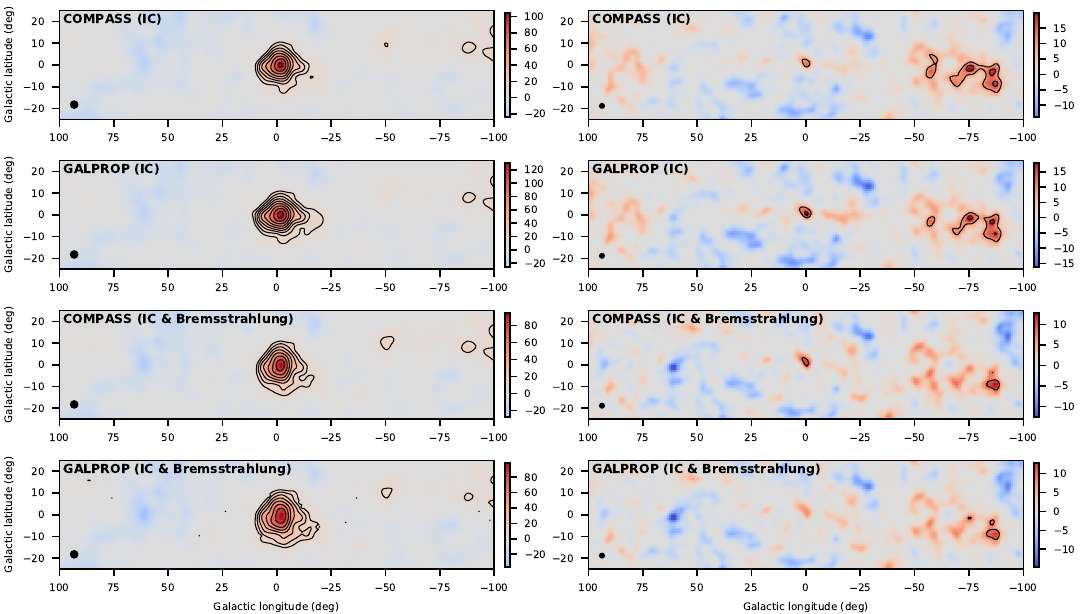}
\caption{
TS maps for the 0.75--3~MeV (left) and 3--30~MeV (right) energy bands for different Galactic 
ridge emission models (see text).
\label{fig:tsmaps-gre}
}
\end{figure*}

To illustrate that the excess emissions are not artefacts of the Galactic ridge emission model
and that they persist under changes of the model, we show in Fig.~\ref{fig:tsmaps-gre} TS maps 
derived using alternative Galactic ridge emission models.
Galactic ridge emission at energies in the MeV range is believed to originate primarily from diffuse 
IC emission of cosmic-ray electrons \citep{strong2000a}; hence, we replaced the 
2D asymmetric Gaussian functions in our reference model with either an IC map employed 
in past COMPTEL data analysis \cite[designated by `COMPASS';][]{youssefi1991} or an IC map 
derived using GALPROP \citep{ackermann2012}.
The resulting TS maps confirm extended $0.75-3$~MeV excess emission towards the Galactic bulge 
and the point-like $3-30$~MeV excess at the Galactic centre.
The $0.75-3$~MeV excess is detected at significance levels exceeding those for the reference 
model and now shows a slightly elongated morphology along the Galactic plane.
This suggests that the IC maps do not fully account for all Galactic ridge emission, which
is confirmed by the presence of significant residual Galactic disc emission in the $3-30$~MeV maps
at negative Galactic longitudes.
The IC maps therefore do not accurately describe the observed Galactic ridge emission, which was 
the principal motivation for employing 2D asymmetric Gaussian functions in our reference model to 
better describe its morphology.

Past COMPTEL data analyses partially compensated for these inadequacies by incorporating
a map of Bremsstrahlung emission into the Galactic ridge emission model, although it remains unclear 
whether the Bremsstrahlung process actually contributes to the observed emission or instead mimics a 
population of unresolved point sources \citep{bloemen1999,strong2000}.
Including Bremsstrahlung maps in the analysis 
\citep[using maps derived for COMPTEL or using GALPROP;][]{strong1993,ackermann2012}
results in TS maps with fewer residual features outside the Galactic bulge compared to the maps 
derived using only the IC templates, although they do not achieve the low residual levels obtained with
the reference model.
The $0.75-3$~MeV excess is confirmed, further demonstrating that its existence does not depend 
on the choice of  Galactic ridge emission model.
The significance of the $3-30$~MeV Galactic centre excess is more sensitive to the choice of
Galactic ridge emission model, falling below $3\sigma$ when a combination of GALPROP IC and 
Bremsstrahlung maps is used.
This decrease is related to a peak in the Bremsstrahlung map towards the Galactic centre, which
reflects the high ISM density in the central molecular zone (CMZ).
The flux in the $3-30$~MeV Galactic centre excess may therefore be overestimated in our
reference model.

\subsection{Characterisation of excess emissions}

To characterise the excess emissions, we added a 2D Gaussian spatial model with a 
power-law spectrum to the reference model and fitted its free parameters (position, extension, 
power-law prefactor, and index) jointly with all free parameters of the reference model.
This resulted in a full width at half maximum (FWHM) of $7.8\dg\pm1.8\dg$ for the emission in the
0.75--3~MeV energy band, with a centroid at $l=-1.4\dg\pm0.8\dg$ and $b=-0.2\dg\pm0.9\dg$, 
slightly offset from the Galactic centre towards negative longitudes. 
The quoted uncertainties are statistical and correspond to a confidence level of 68\%.
Notably, these parameters correspond closely to those of the narrow bulge component of
the 511~keV line emission model of \cite{skinner2014}, which has a FWHM of $7.5\dg$ and
is centred on $l=-1.15\dg$ and $b=-0.25\dg$ (see also below).
The model has a fitted $0.75-3$~MeV flux of $(21.2\pm3.7) \times 10^{-5}$ \pflux\ and a TS value of 
65.5, corresponding to a detection significance of $8.1\sigma$.
Using more complex models, such as elliptical Gaussians with fixed or free position angle, did not 
lead to significant fit improvements.
For the 3--30~MeV energy band, the emission is unresolved, with a FWHM $<3.6\dg$ 
(95\% confidence level) and a centroid at $l=-0.1\dg\pm0.6\dg$ and $b=0.7\dg\pm0.6\dg$.
The model has a fitted $3-30$~MeV flux of $(3.6\pm0.9) \times 10^{-5}$ \pflux\ and a 
TS value of 26.9, corresponding to a detection significance of $5.2\sigma$.

To search for time variability of the excess emissions, we divided the full observing period into four 
equally long time intervals and fitted the excess emission for each of the intervals by fixing the
spatial model parameters to the values that we obtained above.
The results of this analysis are summarised in Table \ref{table:flux}, along with the fluxes derived
for the full observing period.
The fluxes in each time interval are compatible within uncertainties with those for the full period, revealing 
no time variability in the excess emissions.

\begin{table}[t!]
\caption{Fitted excess fluxes, in units of $10^{-5}$ \pflux, for different time intervals.}
\label{table:flux}
\centering
\begin{tabular}{c c c}
\hline\hline
MJD & $0.75-3$~MeV & $3-30$~MeV \\
\hline
Full observing period & $21.2\pm3.7$ & $3.6\pm0.9$ \\
48393 -- 48926 & $30.9\pm6.3$ & $2.5\pm3.9$ \\
48926 -- 49459 & $20.0\pm7.4$ & $4.6\pm1.4$ \\
49459 -- 49992 & $26.1\pm5.3$ & $3.7\pm1.5$ \\
49992 -- 50526 & $17.3\pm7.1$ & $4.4\pm1.6$ \\
\hline
\end{tabular}
\end{table}

\subsection{Comparison with the morphology of the 511~keV line emission}

Motivated by the morphological similarity between the extended $0.75-3$~MeV emission and
the 511~keV line emission from the Galactic bulge, we compared the COMPTEL data in both
energy bands to a model of the Galactic bulge 511~keV line emission derived by
\cite{skinner2014} from ten years of INTEGRAL/SPI data.
This model is the most detailed developed to date and is comprised of three components that include
a point source located at the position of \sgra\ and two symmetric Gaussians modelling the narrow 
and broad bulge emission components.
While the broad component is centred on the Galactic centre, the narrow component is offset from
the Galactic centre towards $l=-1.15\dg$ and $b=-0.25\dg$.
Table \ref{tab:511keV-fit} presents the TS values and fluxes obtained 
for the three model components in both energy bands.
For components that were not detected, 95\% confidence-level upper flux limits were derived.

\begin{table}[t!]
\caption{TS values and fitted fluxes, in units of $10^{-5}$ \pflux, for the three-component
model of \cite{skinner2014} of the Galactic bulge 511~keV line emission.}
\label{tab:511keV-fit}
\centering
\begin{tabular}{c c c c c}
\hline\hline
Component & \multicolumn{2}{c}{$0.75-3$~MeV} & \multicolumn{2}{c}{$3-30$~MeV} \\
& TS & Flux & TS & Flux \\
\hline
Core & $0$ & $<6$ & $12.9$ & $3.7\pm1.7$ \\
Narrow bulge & $9.6$ & $18\pm8$ & $0$ & $<1.8$ \\
Broad bulge & $0.3$ & $4\pm7$ & $0$ & $<3.7$ \\
\hline
\end{tabular}
\end{table}

The log-likelihood value of the fits in both energy bands were comparable to those
obtained by fitting a simple 2D Gaussian model, implying that the model of 
\cite{skinner2014} contains all components needed to explain the excess emissions seen by 
COMPTEL.
In the $0.75-3$~MeV energy band, the `Core' component of \cite{skinner2014} was not detected,
and most of the flux was attributed to the `Narrow bulge' component.
This contrasts the situation for the 511~keV line emission, where \cite{skinner2014} found that
$\sim70\%$ of the flux is attributed to the `Broad bulge' component, implying that the 
$0.75-3$~MeV excess emission is less extended compared to the 511~keV line emission.
In the $3-30$~MeV energy band, all emission is attributed to the `Core' component of \cite{skinner2014},
while no emission is attributed to the bulge components, for which only upper limits could be
derived.
This is consistent with Figs.~\ref{fig:tsmaps} and \ref{fig:tsmaps-gre}, which show no indication of
extended emission in the $3-30$~MeV energy band, confirming the drastic change in emission
morphology at approximately $3$~MeV.

\subsection{Spectral energy distributions}

To advance our understanding of the nature of the excess emissions, we derived spectral energy
distributions (SEDs) for extended and point-like components by performing model fits over the full
$0.75-30$~MeV COMPTEL energy range.
To separate the SED of the extended component from that of the point-like component, we jointly 
fitted two models on top of the reference model: a 2D symmetric Gaussian shape to 
model the extended low-energy emission (hereafter the `Bulge component') and a point source 
to model the point-like high-energy emission (hereafter the `Core component').
We fixed the spatial parameters of both components to the values found in the analysis of the
$0.75-3$~MeV and $3-30$~MeV energy bands.
The spectra of both components were modelled using so-called `bin functions', where piecewise
power laws with fixed spectral index $\Gamma=2$ and free intensity were logarithmically spaced
over the full energy range.
In this way, the number of spectral points in the SED becomes independent of the number of
energy bins used for the data.
We chose eight spectral points for the Bulge component and three spectral points for the Core component 
to capture the spectral evolutions as well as possible while assuring signal detection over as broad an 
energy range as possible.
In addition, we performed fits where the bin functions were replaced by analytical spectral laws,
such as power laws, exponentially cut-off power laws, and log-parabola spectra, to study the spectral
evolution of the excess emissions independently of any spectral binning.
We translated uncertainties in the spectral parameters into uncertainty bands that account for the covariance 
between the spectral parameters of the Bulge and Core components.

\begin{figure}[!t]
\centering
\includegraphics[width=\columnwidth]{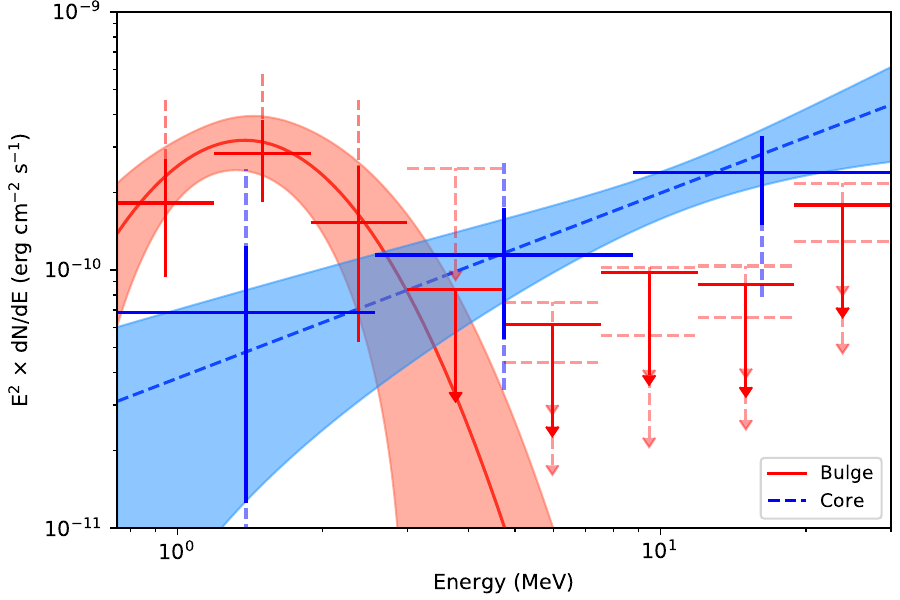}
\caption{Spectral energy distributions and fitted spectral laws for the Bulge and Core components.
Systematic uncertainties related to different modelling of the Galactic ridge emission are shown as 
light dashed vertical bars.
For upper limits,  we show the smallest and largest values obtained under variation of
the Galactic ridge emission model as faint dashed arrows.}
\label{fig:iased}
\end{figure}

The resulting SEDs and fitted analytical spectra for the Bulge and Core components are shown in 
Fig.~\ref{fig:iased}.
As expected from the excess emission maps, the Bulge component dominates the emission at low 
energies.
Above $\sim$2 MeV the energy flux starts to drop, and above $\sim$3 MeV only upper flux limits 
could be derived, consistent with the absence of extended emission towards the Galactic bulge 
region in the $3-30$~MeV excess emission map.
The Bulge component is well fitted by either an exponentially cut-off power law or a log-parabola 
spectrum.
For the latter, we derived $0.75-30$~MeV photon and energy fluxes of 
(1.5$\pm$0.4) $\times$ 10$^{-4}$ \pflux\ and
(3.4$\pm$0.9) $\times$ 10$^{-10}$ \eflux.
For a source situated at the distance of the Galactic centre \citep[8.2 kpc;][]{gravity2019}, the flux
corresponds to a luminosity of (2.7$\pm$0.7) $\times$ 10$^{36}$ \elum, which is approximately 10\%
of the annihilation luminosity from thermalised positrons \citep{churazov2020}.

The Core component shows a distinctively different SED, with a steady rise of the energy flux over the 
COMPTEL energy range, compatible with a simple power law.
We derived $0.75-30$~MeV photon and energy fluxes of 
(5.9$\pm$2.4) $\times$ 10$^{-5}$ \pflux\ and
(5.7$\pm$1.2) $\times$ 10$^{-10}$ \eflux, corresponding to a 
luminosity of (4.5$\pm$1.0) $\times$ 10$^{36}$ \elum\ for a source at the distance of the Galactic 
centre.

\subsection{Comparison to expected positron in-flight annihilation spectra}

As the morphology of the Bulge component suggests positron in-flight annihilation as the possible
origin of the emission, we investigated whether the spectrum is also compatible with such a
scenario and whether conclusions can be drawn about the underlying positron sources.
For this purpose, we modelled the \gray\ emission from a population of energetic positrons interacting 
with electrons in the ISM, following the prescriptions of \cite{beacom2006}, \cite{sizun2006}, and 
\cite{prantzos2011}.
While these prescriptions are limited to the case of stationary injection of mono-energetic positrons 
into the ISM, we extended the formalism to arbitrary injection spectra.
Following \cite{beacom2006}, we normalised the resulting \gray\ intensities to be in units of equivalent 
511~keV line flux $\Phi_{511}$, using the observed positronium fraction of $f_{\rm Ps}$ = 0.967 
\citep{jean2006}, so that when fitting the models to the COMPTEL data, the model normalisation 
is given in units of $\Phi_{511}$ (see Appendix \ref{sec:modelling}).

Figure \ref{fig:expected-spectra} shows model spectra for the Bulge and Core components resulting 
from a joint fit of both components to the data over the $0.75-30$~MeV energy band.
The Bulge component was modelled using a 2D Gaussian spatial model with parameters
fixed to the values determined from the $0.75-3$~MeV energy band, combined with a spectral positron 
in-flight annihilation model for which $\Phi_{511}$ was fitted.
Additional free parameters include the injection energy, $E_p$, for mono-energetic models (panels a
and b), the maximum injection energy, $E_{max}$, and the spectral index, $\Gamma$, for the power-law 
particle injection spectra (panels c and d).
In all fits, the Core component was modelled by a point source located at the position obtained 
in the $3-30$~MeV energy band with a power-law spectrum.

\begin{figure*}[!t]
\centering
\includegraphics[width=\textwidth]{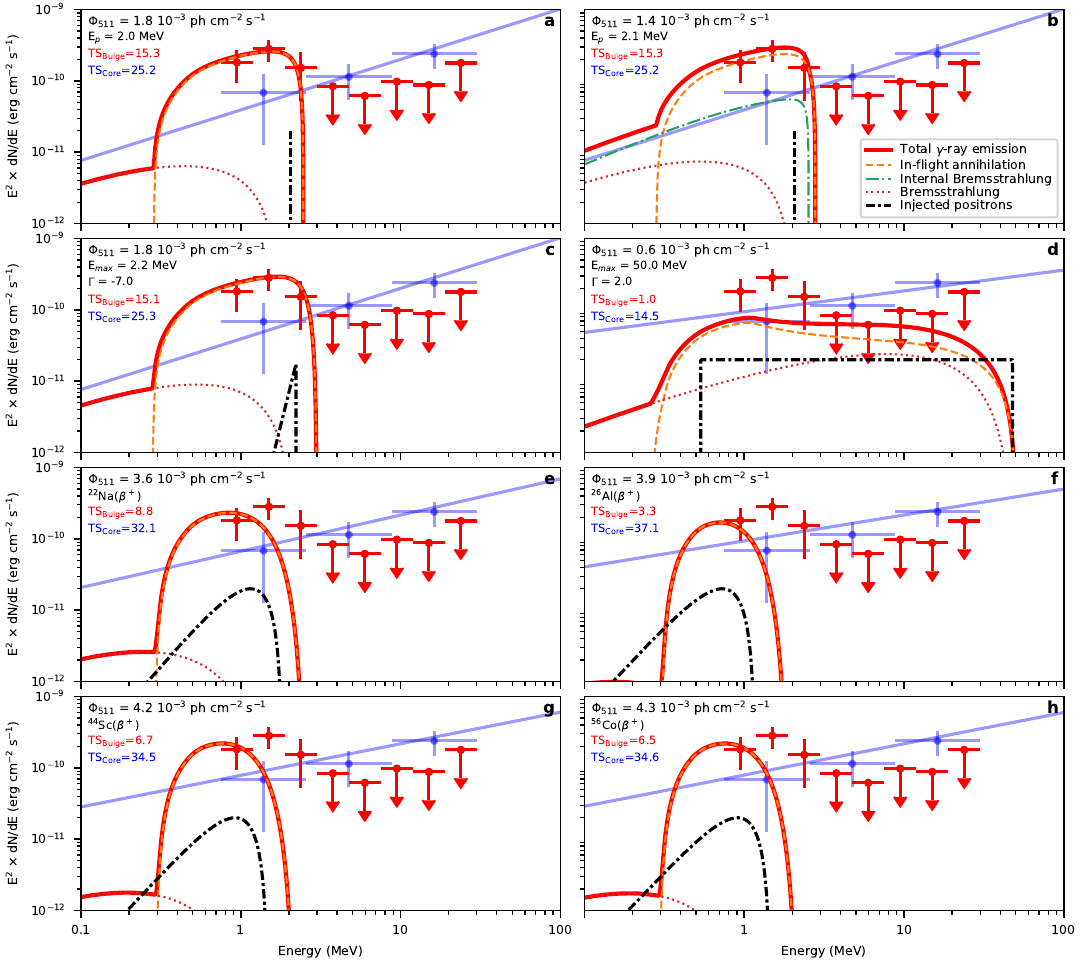}
\caption{Expected \gray\ spectra for positrons in flight (red lines), including contributions from
in-flight annihilation (dashed orange lines), Bremsstrahlung (dotted brown lines), and internal
Bremsstrahlung (dash-dotted green  line).
Positron injection spectra are shown as dash-dotted black lines.
Model spectra are shown for stationary injection of mono-energetic positrons (a), dark matter 
annihilation (b), stationary injection of positrons obeying a power-law spectrum of slope $\Gamma$ 
and maximum energy $E_{\rm max}$ (c, d), and $\beta^+$ decays of radioactive isotopes (e--h).
The power-law spectra of the jointly fitted Core component are shown as light blue lines.
The SEDs of the Bulge and Core components, replicated from Fig.~\ref{fig:iased}, are shown as red and light 
blue data points for visual comparison.
Fitted parameters for the Bulge component and TS values for the Bulge and Core 
components are given in each panel.}
\label{fig:expected-spectra}
\end{figure*}

The Bulge component is best fitted by a spectral model that assumes the stationary injection of 
mono-energetic positrons with kinetic energies of $2.0\pm0.1$~MeV into the ISM (panel a).
The injection energy increases slightly to $2.1\pm0.1$~MeV in the case of dark matter annihilation, 
owing to the additional internal Bremsstrahlung emission component (panel b).
Both models result in a TS value of 15.3 for the Bulge component, corresponding to a detection 
significance of $3.9\sigma$ for the Bulge component on top of a model comprising only the
Core component.
An equally good fit is obtained by assuming the injection of positrons with a power-law spectrum 
$N(E_p) \propto E_p^{-\Gamma}$ of kinetic energies that is cut off at a maximum energy of 
$2.2\pm0.7$~MeV, resulting in a TS value of 15.1 for the Bulge component (panel c).
The fitted power-law slope of $\Gamma\sim-7$ is extremely hard, resulting in a spectrum that is
heavily peaked towards the maximum kinetic energy, implying that it is quasi mono-energetic.
Forcing the slope of the injection spectrum to $\Gamma=2$ and setting $E_{max}=50$~MeV 
results in a marginal TS value of 1.0 for the Bulge component (panel d), demonstrating that the
injection of a broad-band spectrum of positrons into the ISM to explain the Bulge component is 
strongly excluded by the COMPTEL data.
Finally, fitting the expected spectra for positrons produced by the $\beta^+$ decays of 
$^{22}$Na (panel e),
$^{26}$Al (panel f),
$^{44}$Sc (panel g), and
$^{56}$Co (panel h) results in TS values below the $3\sigma$ significance level for the
Bulge component, essentially because these spectra cut off at energies below the observed
cut-off of the Bulge component.

\section{Discussion}

\subsection{Possible origin of the Bulge component}

The resemblance of the morphology of the Bulge component to that of the 511~keV line bulge emission
is striking.
Both emissions are extended, with comparable sizes, and exhibit an emission centroid that is slightly 
offset from the Galactic centre towards negative longitudes.
The characteristics of this emission morphology are unique, having no counterpart in the gas components 
or stellar populations of the Galaxy \citep{prantzos2011}.
Consequently, we suggest that the Bulge component most plausibly originates from positron in-flight 
annihilation.

After production, positrons do not annihilate immediately owing to the relatively low densities of the 
ISM \citep{prantzos2011}.
Survival times may range from $\sim10^3$ years in the dense regions around the Galactic centre
to several times $10^5$ years in the Galactic disc, during which time positrons will propagate away 
from their sources.
As positron in-flight annihilation occurs before positron thermalisation, the in-flight annihilation emission 
should have a more compact morphology compared with the 511~keV line emission.
However, because in-flight annihilation is most probable after positrons have lost
much of their initial energy, it tends to occur after positrons have travelled substantial distances. 
Hence, the extension of the emissions may not differ significantly.
Comparison of the morphology of the Bulge component with morphology models of 511~keV line
emission (cf.~Table \ref{tab:511keV-fit}) suggests that the Bulge component is more compact than 
the 511~keV line emission, in line with expectations if the Bulge component originates from
positron in-flight annihilation.

The SED of the Bulge component is difficult to explain with cosmic-ray-induced IC or 
Bremsstrahlung emissions, as these have broad-band spectra spanning many decades in energy and do 
not show the observed spectral cut-off \citep{strong2000a}.
As illustrated in Fig.~\ref{fig:expected-spectra}, the Bulge component is best fitted by a 
quasi mono-energetic injection of positrons with kinetic energies of $\sim$2 MeV into the ISM.
This excludes radioactive $\beta^+$ decays as noticeable contributors to
the Bulge component, since their positron spectra are limited to sub-MeV energies, producing \gray\ 
spectra that cut off below $\sim$1~MeV.
Some reacceleration of $\beta^+$ decay positrons would be required to increase the injection energies 
to the observed values; however, it is unclear how such a scenario would produce the quasi mono-energetic 
energy distribution observed for the Bulge component.
Scenarios injecting highly relativistic positrons into the ISM -- such as cosmic-ray secondaries produced 
from $\pi^+$ decays, normal pulsars, millisecond pulsars, and magnetars -- are also excluded as positron 
sources of the Bulge component, since they would lead to a positron in-flight annihilation spectrum 
extending to much higher energies.

One proposed source producing mono-energetic positrons with the required kinetic
energies is the annihilation of light dark matter particles \citep{boehm2004}.
If such particles exist, their annihilation would result in MeV \gray\ emission dominated by positron in-flight 
annihilation, with additional smaller contributions from internal and external Bremsstrahlung 
\citep{beacom2006,sizun2006,boehm2006}.
Fitting the expected spectra to the COMPTEL data suggests a dark matter particle mass of $\sim$3~MeV
and implies a 511~keV line flux of (1.5$\pm$0.3) $\times$ 10$^{-3}$ \pflux.
For a source at the distance of the Galactic centre, this flux corresponds to a positron production 
rate of (2.3$\pm$0.5) $\times$ 10$^{43}$ \psec.
This is at the upper end of the range of positron annihilation rates inferred from INTEGRAL/SPI 
511~keV line data for the bulge component, but compatible with estimates for the entire Galaxy, including the 
bulge and a putative disc component \citep{churazov2020}.
This may indicate that some positrons escape from the bulge into the disc region, where they
thermalise and finally annihilate, producing the 511~keV line disc emission. 

The offsets of the centroids of the bulge emissions towards negative longitudes, however, present a challenge 
for the dark matter scenario.
Offsets could, in principle, be explained by ISM overdensities that enhance positron annihilation, yet the 
ISM distribution towards the Galactic centre is offset towards positive, not negative, 
longitudes \citep{ferriere2008}.
This excludes the ISM distribution as an explanation and suggests that the offset is instead related to the 
positron injection site.
While it cannot be excluded that the barycentre of the dark matter particles is offset from the dynamical 
centre of the Galaxy \citep{kuhlen2013}, tidal forces produced by baryons, which dominate the gravitational
potential near the Galactic centre, should remove such an offset within $\sim$10 million 
years \citep{gorbonov2013}.

Another issue with the dark matter hypothesis is the high ISM density in the CMZ around the Galactic 
centre \citep{ferriere2008}, which implies that any positron injected into that zone will annihilate within a 
few thousand years and will not propagate more than a few tens of parsecs at most.
Therefore, positron sources located within the CMZ should lead to strongly peaked annihilation
emission towards the Galactic centre, with an angular extent less than the
$\sim2\dg$ size of the CMZ \citep{alexis2014}.
Since the Bulge component does not show such a compact source (cf.~Table \ref{tab:511keV-fit}), it is 
unlikely that the source giving rise to the extended Bulge component is located in the immediate vicinity
of the Galactic centre.
This excludes, in particular, \object{\sgra} as the main positron source for the Bulge component.

A candidate positron source seen near the Galactic centre, but located at the edge or possibly 
even outside the CMZ, is the microquasar \object{1E~1740.7--2942} \citep{luque2015,tetarenko2020}. 
Intriguingly, its position coincides with the centroid of the 511~keV line bulge emission \citep{skinner2014}
as well as with the centroid of the Bulge component.
\object{1E~1740.7--2942} has been dubbed the `Great Annihilator' owing to past reports of transient positron 
annihilation features in its spectrum \citep{bouchet1991,sunyaev1991}.
Although these observations have been questioned \citep{prantzos2011}, \object{1E~1740.7--2942} remains 
a plausible source of positrons, since microquasars may produce positrons in the inner region of the 
accretion disc or hot corona, which are then channelled into the ISM through their jets \citep{guessoum2006}.
While microquasars are variable sources, their variability occurs on timescales much shorter
than the positron lifetime in the ISM. Hence, regardless of whether positrons are produced
during outbursts or rather continuously, microquasars should act as steady sources of positron 
annihilation.
The time-averaged jet power of \object{1E~1740.7--2942} is estimated to be $(0.7-3.5) \times 10^{37}$ 
\elum\,\,\citep{tetarenko2020}, which is sufficient to produce the observed positron annihilation 
luminosity of the Bulge component if the jet is composed of a pure pair plasma.
The plasma's bulk motion will inject positrons with relatively constant kinetic energies into the 
ISM \citep{heinz2002}, which would explain the quasi mono-energetic positron spectrum that we inferred 
for the Bulge component.
Bulk Lorentz factors of microquasar jets are uncertain, but values of up to $\Gamma\sim$~4 are
discussed in the literature \citep{luque2015,zdziarski2022}.
This would correspond to kinetic energies of $\sim$1.5 MeV, which is close to the value inferred for
the Bulge component.
After injection, positrons propagate along magnetic field lines away from the source while slowing 
down due to scattering on gas particles and possibly plasma waves \citep{jean2009}, giving 
rise to extended positron annihilation emission whose morphology depends on the magnetic field 
configuration and the density and dynamics of the ISM \citep{alexis2014}.
While some positrons may annihilate in the CMZ, the strong turbulence in the inner Galaxy, 
combined with gas outflows \citep{sarkar2024}, may transport positrons away from the 
source, leading to the observed extended emission feature \citep{panther2018}.

The obvious question in this scenario is why no positron annihilation emission is observed
around other microquasars.
The object \object{1E~1740.7--2942} is unique among microquasars, remaining persistently in the 
bright outburst state and accreting near peak luminosity most of the time, which are properties shared 
only with \object{GRS~1758--258} \citep{tetarenko2016}.
However \object{GRS~1758--258} has a lower jet power \citep{tetarenko2020} and is located further away from the 
Galactic centre, placing it in a less dense region of the ISM.
Lower ISM densities imply larger positron propagation distances and higher escape fractions,
reducing positron annihilation emission intensities.
The combination of a powerful positron source located in a dense ISM region could explain why
\object{1E~1740.7--2942} is the only currently detectable microquasar source of positrons.

\subsection{Possible origin of the Core component}

\begin{figure}[!t]
\centering
\includegraphics[width=\columnwidth]{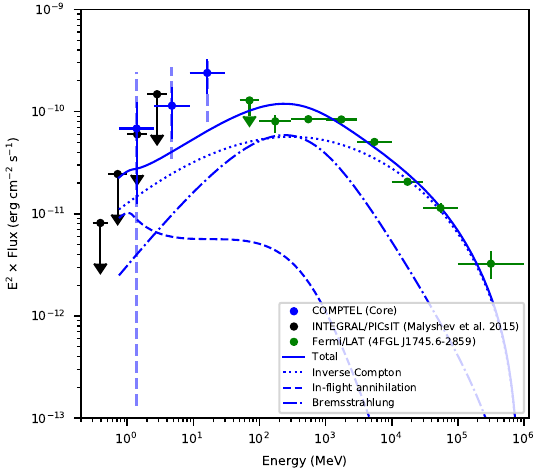}
\caption{Modelling the SED of the Core component.
The Core component SED is shown together with the SED of the Fermi-LAT source 
\object{4FGL~J1745.6--2859} \citep{4fgl} and upper limits from INTEGRAL/PICsIT \citep{malyshev2015} (black). 
A best-fitting model for a pair plasma injected by \sgra\  is shown as a solid blue line.
Contributions to the model from IC (dotted line), positron in-flight annihilation (dashed line),
and Bremsstrahlung (dash-dotted line) emissions are also shown.}
\label{fig:core}
\end{figure}

The SED of the Core component differs significantly from that of the Bulge component 
(cf.~Fig.~\ref{fig:iased}), suggesting distinct origins.
It is nevertheless intriguing that both the 511~keV line and MeV continuum emissions exhibit a 
compact component at the Galactic centre \citep{skinner2014,siegert2016}, raising the question 
of whether the Core component is also related to positron annihilation.
A central component could, in principle, be produced by the enhanced gas density in the CMZ, 
leading to enhanced positron annihilation emission; however, in that scenario, the SEDs of the 
Core and Bulge components would be similar.

The missing link is likely the compact source in the GeV energy range, \object{4FGL~J1745.6--2859}, 
observed by Fermi-LAT at the Galactic centre \citep{malyshev2015,4fgl,cafardo2021}. 
This is by far the brightest Fermi-LAT source in the field of the Core component.
As illustrated in Fig.~\ref{fig:core}, the SEDs of the Core component and \object{4FGL~J1745.6--2859} 
connect rather well, making the Fermi-LAT source the most plausible high-energy counterpart of 
the Core component.
While the $8.8-30$~MeV flux of the Core component is slightly in excess of a smooth spectral 
connection, the modelling of the Galactic ridge emission introduces some systematic uncertainties 
in the flux levels. 
This also applies for the Fermi-LAT low-energy fluxes \citep{malyshev2015,4fgl,cafardo2021}.
Within these uncertainties, the connection between the two SEDs is satisfactory.

\object{4FGL~J1745.6--2859} is a non-variable \gray\ source of unknown nature. 
However, the production of relativistic hadrons or leptons by \sgra\ is generally invoked to explain its GeV 
emission \citep{cafardo2021}.
Fig.~\ref{fig:core} shows the \gray\ spectrum arising from the continuous injection of a 
relativistic pair plasma into the ISM at the Galactic centre, with a particle spectrum slope $\Gamma$ 
of 2.1 and maximum energy of 1~TeV.
This model satisfactorily explains the observed SEDs of the Core component and \object{4FGL~J1745.6--2859}, 
while simultaneously producing a 511~keV line emission of $10^{-4}$ \pflux. 
This flux corresponds to the observed flux for the central 511~keV line emission component 
\citep{skinner2014,siegert2016}.
In such a model, the \gray\ emission would primarily be IC emission due to the high 
radiation field at the Galactic centre \citep{kusunose2012}, with contributions from positron in-flight 
annihilation and Bremsstrahlung at low energies.
The high radiation field and large ISM density at the Galactic centre imply that positrons will thermalise 
quickly and propagate less, explaining the compact appearance of the source.
Once thermalised, positron annihilation will lead to 511~keV line emission which produces the central
component observed in the INTEGRAL/SPI data \citep{skinner2014,siegert2016}.

\section{Conclusions}

By analysing archival data from the COMPTEL telescope we establish the presence of
MeV excess emission on top of the Galactic ridge \gray\ emission towards the Galactic centre region.
The morphology of this excess emission changes markedly within the COMPTEL $0.75-30$~MeV
energy range, switching from an extended emission morphology below $\sim3$~MeV to a point-like
appearance above that energy.
A spectral analysis reveals the presence of two distinct excess emission components towards the 
Galactic centre.

We plausibly identify the first component to originate from positron in-flight annihilation and reveal
that its spectrum matches a scenario in which Galactic bulge positrons are produced
by dark matter annihilation \citep{beacom2006,sizun2006,boehm2006}.
If this scenario applies, we estimate the rest mass of the dark matter particles to be $\sim3$~MeV.
The only other candidate source, besides dark matter annihilation, that satisfies the observations is 
positron injection through the jets of the microquasar \object{1E~1740.7--2942}; however, it remains 
unclear whether that scenario can satisfactorily explain the extended morphology of the emission.
Thus, the precise origin of positrons in the Galactic bulge remains unknown, but our 
observations exclude many sources previously proposed \citep{prantzos2011}.
In particular, $\beta^+$ decays from radioactive isotopes and sources producing highly relativistic 
positrons, such as cosmic-ray secondaries produced from $\pi^+$ decays, normal pulsars, millisecond 
pulsars, and magnetars, can be clearly discarded as primary source of Galactic bulge positrons.

We plausibly identify the second component as the low-energy counterpart of the Fermi-LAT
point source \object{4FGL~J1745.6--2859}, which coincides with the location \object{\sgra}.
While the production of relativistic hadrons or leptons by \object{\sgra} has been invoked as the origin of
the emission \citep{cafardo2021}, the detection of a low-energy counterpart by COMPTEL 
favours a leptonic scenario.
The broad-band spectrum of the source is well explained by the injection of a relativistic pair plasma 
into the ISM, and the high matter density at the Galactic centre would limit plasma propagation and 
explain the point-like appearance of the source.
Interestingly, the pair plasma scenario also explains the point-like component in the
511~keV line emission observed by INTEGRAL/SPI towards the Galactic bulge, attributing it to the 
annihilation of thermalised positrons in the plasma.

Our work illustrates the discovery potential inherent in the exploitation of archival data, 
enabling science advances with modest environmental impact.
We estimate that the computing for this analysis produced about 600 kgCO$_2$e of greenhouse 
gases, which is about 40 times less than the median per-publication emissions associated with the 
analysis of data from an active astronomical observatory \citep{knoedlseder2022b}.
Expanding the exploitation of archival data, while simultaneously reducing the
deployment pace of new facilities, therefore provides a powerful measure to reduce the carbon footprint 
of astrophysical research while continuing to advance the field.

\begin{acknowledgements}
This work has made use of the Python 2D plotting library matplotlib \citep{hunter2007}.
\end{acknowledgements}

\bibliographystyle{aa}
\bibliography{aa56046-25}

\begin{appendix}

\section{Modelling of instrumental background events}
\label{sec:background}

COMPTEL data are dominated by instrumental background events that arise from cosmic-ray particles,
geomagnetically trapped radiation-belt particles, albedo neutrons, and \gray\ photons that
interact with the spacecraft and detector materials \citep{weidenspointner2001}.
Event rates due to celestial \gray\ sources are much lower than instrumental background
event rates, with typical signal-to-noise ratios of a few percent.
Consequently, the accurate modelling of the distribution of instrumental background events in the
three-dimensional COMPTEL data space is crucial for a reliable determination of \gray\
source characteristics from the data.

For our analysis we modelled the instrumental background event distribution using an algorithm that 
we dubbed BGDLIXF, which is an evolution of the BGDLIXE algorithm that was introduced in 
section 3.5.3 of \cite{knoedlseder2022a}.
While BGDLIXE has been shown to provide an accurate model of the instrumental background event 
distribution for individual viewing periods \citep{knoedlseder2022a}, the algorithm does not cope with 
background event rate variations between viewing periods, which is problematic when viewing periods 
are combined for analysis.
Viewing periods have typical durations of two weeks during which the CGRO satellite had a stable 
pointing, and for our analysis we combined 215 viewing periods that span a duration of almost 6
years of observations.
During this period the instrumental background event rates varied substantially \citep{weidenspointner1999},
and ignoring these variations introduces significant features in the TS maps that are attributable to 
inadequacies of the instrumental background model.

To introduce the BGDLIXF algorithm, we start by rewriting the BGDLIXE algorithm as
\begin{equation}
\begin{array}{l}
{\tt DRB}_i(\chi,\psi,\bar{\varphi}) =  \, {\tt DRW}_i(\chi,\psi,\bar{\varphi}) \, \times \\
\displaystyle\,\,\,\,\,\,\,\,\,\,
\frac{\sum_{\chi' \in \{ A_{\chi} \}} \sum_{\psi' \in \{ A_{\psi} \}} \sum_{\bar{\varphi}' \in \{ R_{\bar{\varphi}} \} } {\tt DRE}_i(\chi', \psi', \bar{\varphi}') - {\tt DRM}_i(\chi', \psi', \bar{\varphi}')}
{\sum_{\chi' \in \{ A_{\chi} \}} \sum_{\psi' \in \{ A_{\psi} \}} \sum_{\bar{\varphi}' \in \{ R_{\bar{\varphi}} \} } {\tt DRW}_i(\chi',\psi',\bar{\varphi}')}
\label{eq:supp-bgdlixf}
\end{array}
\end{equation}
where ${\tt DRB}_i(\chi,\psi,\bar{\varphi})$ is the background model for viewing period $i$ 
and ${\tt DRW}_i(\chi,\psi,\bar{\varphi})$ corresponds to the term 
${\tt DRB}_{\rm phinor}(\chi,\psi,\bar{\varphi})$ used in \cite{knoedlseder2022a} that we 
call here the `weighting cube' for viewing period $i$.
Furthermore, ${\tt DRE}_i(\chi,\psi,\bar{\varphi})$ is the data cube of observed events and
${\tt DRM}_i(\chi,\psi,\bar{\varphi})$ is the data cube of modelled events from celestial
\gray\ sources for viewing period $i$.
The variables $\chi$, $\psi$, and $\bar{\varphi}$ are the Compton scatter direction and
scatter angle that span the three-dimensional COMPTEL data space, while $\{ A_{\chi} \}$, 
$\{ A_{\psi} \}$ and $\{ R_{\bar{\varphi}} \}$ correspond to running intervals in each data space 
dimension that are used for averaging \citep{knoedlseder2022a}.
Weighting cubes are first order background models for each viewing period based on the
geometry functions ${\tt DRG}_i(\chi, \psi, \bar{\varphi})$ that are normalised so that they 
reproduce the observed number of background events and are given by
\begin{eqnarray}
{\tt DRW}_i(\chi,\psi,\bar{\varphi}) & = &
{\tt DRG}_i(\chi, \psi, \bar{\varphi}) \, \Omega(\chi, \psi) \, \times \nonumber\\
&&\displaystyle
\frac{\sum_{\chi',\psi'} {\tt DRE}_i(\chi', \psi', \bar{\varphi}) - {\tt DRM}_i(\chi', \psi', \bar{\varphi}))}
{\sum_{\chi',\psi'} {\tt DRG}_i(\chi', \psi', \bar{\varphi}) \, \Omega(\chi', \psi')}
\label{eq:supp-bgdlixfconst}
\end{eqnarray}
Here the summations are taken over all Compton scatter angles $\chi'$ and $\psi'$, and the
geometry functions are multiplied with the solid angle $\Omega(\chi, \psi)$ of the data space 
cells to convert interaction probabilities to absolute expected event rates.
Note that Eq.~(\ref{eq:supp-bgdlixfconst}) corresponds to the background model produced
by the PHINOR algorithm that is described in section 3.5.1 of \cite{knoedlseder2022a}.

In the original BGDLIXE algorithm, Eq.~(\ref{eq:supp-bgdlixf}) was also applied to the combined ${\tt DRE}$, 
${\tt DRM}$ and ${\tt DRG}$ data cubes produced by the ${\tt comobsadd}$ script, where the combined ${\tt DRG}$ 
is the sum of the exposure weighted geometry functions of the individual viewing periods 
\citep[see Eq.~(21) of][]{knoedlseder2022a}.
This formulation implicitly assumes that the instrumental background rate is the same for
all viewing periods, and that the number of instrumental background events is directly proportional 
to exposure time.
In reality, however, the instrumental background rate evolves over the CGRO mission due to the
varying radiation environment along the orbits and the build-up of radioactive isotopes within the
satellite \citep{weidenspointner1999}.
These variations are particularly important at low energies, and ignoring them when combining viewing 
periods leads to important biases in the model fits and significant background residuals in sky maps.

Introducing a weighting cube ${\tt DRW}_i(\chi, \psi, \bar{\varphi})$ for each viewing period circumvents 
these problems and enables modelling of temporal variations in the background rate when combining 
viewing periods.
We extended ${\tt comobsadd}$ that now also computes the combined weighting cube
\begin{equation}
{\tt DRW}(\chi, \psi, \bar{\varphi}) = \sum_i {\tt DRW}_i(\chi, \psi, \bar{\varphi})
\end{equation}
using the same formula that is applied for combining the observed events.
Since each ${\tt DRW}_i(\chi, \psi, \bar{\varphi})$ is normalised to the observed number of background 
events, the combined ${\tt DRW}(\chi, \psi, \bar{\varphi})$ correctly takes into account variations in the 
background rate between viewing periods.

To incorporate also background rate variations within the time span of a viewing period, as experienced 
due to the varying radiation environment along the orbits of the CGRO satellite \citep{weidenspointner1999},
we replaced in Eq.~(\ref{eq:supp-bgdlixfconst}) the term ${\tt DRG}_i$ by the more elaborate computation
\begin{eqnarray}
{\tt DRW}_i(\chi,\psi,\bar{\varphi}) & = &
\Omega(\chi, \psi) \sum_k R_k(\bar{\varphi}) \, \tilde{G}_k(\chi, \psi, \bar{\varphi}) \, \times \nonumber\\
&&\frac{\sum_{\chi',\psi'} {\tt DRE}_i(\chi', \psi', \bar{\varphi}) - {\tt DRM}_i(\chi', \psi', \bar{\varphi})}
{\sum_{\chi',\psi',k} R_k(\bar{\varphi}) \, \tilde{G}_k(\chi, \psi, \bar{\varphi}) \, \Omega(\chi', \psi')}
\label{eq:supp-bgdlixfrate}
\end{eqnarray}
where
$R_k(\bar{\varphi})$ is a $\bar{\varphi}$-dependent event rate and $\tilde{G}_k(\chi, \psi, \bar{\varphi})$ 
is the geometry function for each superpacket $k$, which is a bunch of 8 spacecraft telemetry packets 
of 16.384 s duration.
To understand Eq.~(\ref{eq:supp-bgdlixfrate}) we recall that
\begin{equation}
{\tt DRG}_i(\chi,\psi,\bar{\varphi}) = \frac{1}{N_i} \sum_k \tilde{G}_k(\chi, \psi, \bar{\varphi})
\end{equation}
where $N_i$ is the number of superpackets in viewing period $i$ 
\citep[see Eq.~(F.1) in][]{knoedlseder2022a}.
Hence the computation of ${\tt DRG}_i(\chi,\psi,\bar{\varphi})$ assumes that each superpacket has the 
same weight (as needed for the computation of the instrument response), while 
Eq.~(\ref{eq:supp-bgdlixfrate}) weights the superpackets with $R_k(\bar{\varphi})$.
We determined $R_k(\bar{\varphi})$ from the data themselves by counting the number of events for 
each $\bar{\varphi}$ interval within a sliding window of 5 minutes centred on each superpacket $k$.
Taking into account the event rate variations within a viewing period slightly reduces backgound
residuals in TS maps and improves the model fit as measured by the maximum log-likelihood value
with respect to a background model based on Eq.~(\ref{eq:supp-bgdlixfconst}).
For practical purposes, the computation of ${\tt DRW}_i(\chi,\psi,\bar{\varphi})$ was implemented in the 
${\tt comobsbin}$ script, where Eq.~(\ref{eq:supp-bgdlixfconst}) is used when the 
${\tt drwmethod=CONST}$ option is selected, while Eq.~(\ref{eq:supp-bgdlixfrate}) is used for the 
${\tt drwmethod=PHIBAR}$ option.

Finally, the predicted events ${\tt DRM}_i(\chi, \psi, \bar{\varphi})$ attributed to celestial \gray\
emission are only known once a model has been fitted to the data, introducing a circular dependency
between models of celestial emission and instrumental background.
To solve this dependency we used an iterative approach, starting from an initial fit where we set 
${\tt DRM}_i(\chi, \psi, \bar{\varphi})$ to zero.
As a consequence, the background rate, and hence ${\tt DRW}_i(\chi,\psi,\bar{\varphi})$, will be slightly 
overestimated in the presence of a \gray\ signal in this first fit.
In subsequent iterations, the fitted celestial model will be used to recompute 
${\tt DRW}_i(\chi,\psi,\bar{\varphi})$, a procedure that converges after a few iterations.
We note that this iterative procedure, together with the BGDLIXF algorithm, corresponds to the 
${\tt SRCFIX}$ method that was introduced near the end of the COMPTEL mission in the COMPASS
data analysis system, and that is used for most COMPTEL science analyses since then \citep{bloemen1999}.
The only aspect that is new with respect to that method is the incorporation of background rate
variations within a viewing period (cf.~Eq.~\ref{eq:supp-bgdlixfrate}), improving slightly the background 
models with respect to the ${\tt SRCFIX}$ approach.

\section{Modelling of known point sources}
\label{sec:sources}

In our analysis we always fitted ten point sources of continuum \gray\ emission to the data that were 
known from previous work to emit in the COMPTEL energy range 
\citep{schoenfelder2000,zhang2002,collmar2006}.
The locations of the sources were kept fixed at their nominal position, and all sources were fitted using
power-law spectral models.
The sources and their best-fitting spectral parameters that were obtained by fitting the data over the 
0.75--30 MeV energy band are summarised in Table \ref{tab:ptsrc}.
We also searched for \gray\ emission from additional sources that were reported in the literature 
(GRO~J1036--55,
3C~454.3,
PSR~J0659+1414,
PSR~J1057--5226,
PSR~J1952+3252,
PSR~J2229+6114,
Geminga,
LSI +61\dg\,303,
Nova Per 1992,
PKS 0528+134,
PKS 1222+216,
PKS 1622--297,
PKS 2230+114,
PMN J1910-2448),
yet we did not detected significant emission from any of these sources in our analysis,
which was restricted in sky and time coverage.

\begin{table*}
\caption{Point sources fitted in the analysis.}
\label{tab:ptsrc}
\centering
\begin{tabular}{l c c c c c c c}
\hline\hline
Source & $l$ & $b$ & TS & k & $\Gamma$ & Photon flux & Energy flux \\
\hline
PKS~1830--210 & 12.17 & -5.71 & 24.4 & 3.2$\pm$0.9 & 2.2$\pm$0.2 & 1.0$\pm$0.4 & 3.8$\pm$1.1 \\
LS~5039 & 16.88 & -1.29 & 85.2 & 7.3$\pm$0.9 & 1.5$\pm$0.1 & 1.3$\pm$0.2 & 10.4$\pm$1.2 \\
Cyg~X-1 & 71.33 & 3.07 & 252.8 & 6.8$\pm$1.0 & 2.5$\pm$0.1 & 2.9$\pm$0.6 & 8.9$\pm$1.4 \\
Vela pulsar & 263.55 & -2.79 & 31.6 & 5.3$\pm$1.0 & 1.5$\pm$0.2 & 1.0$\pm$0.2 & 7.4$\pm$1.4 \\
PKS~0506--612 & 270.55 & -36.07 & 16.4 & 2.0$\pm$1.1 & 2.5$\pm$0.4 & 0.9$\pm$0.6 & 2.7$\pm$1.5 \\
3C~273 & 289.95 & 64.36 & 85.3 & 7.1$\pm$0.8 & 1.8$\pm$0.1 & 1.6$\pm$0.2 & 8.7$\pm$1.0 \\
3C~279 & 305.10 & 57.06 & 32.8 & 4.1$\pm$0.7 & 1.8$\pm$0.2 & 0.9$\pm$0.2 & 5.1$\pm$0.9 \\
Cen~A & 309.52 & 19.42 & 164.8 & 8.1$\pm$1.0 & 2.1$\pm$0.1 & 2.4$\pm$0.4 & 9.7$\pm$1.2 \\
GRO~J1411--64 & 311.50 & -2.50 & 38.3 & 2.2$\pm$1.2 & 2.7$\pm$0.4 & 1.3$\pm$0.9 & 3.4$\pm$2.1 \\
PSR~J1513--5908 & 320.32 & -1.16 & 49.3 & 6.8$\pm$1.0 & 1.6$\pm$0.1 & 1.3$\pm$0.2 & 9.0$\pm$1.3 \\
\hline
\end{tabular}
\tablefoot{
$l$ and $b$ specify the Galactic longitudes and latitudes of the point sources in degrees, TS gives the 
test statistic of the source detection, $k$ are the power-law prefactors, and $\Gamma$ are the spectral 
indices.
Power-law prefactors are in units of $10^{-6}$ ph cm$^{-2}$ s$^{-1}$ MeV$^{-1}$ and specified for a 
pivot energy of 4.5 MeV.
Photon and energy fluxes are specified for the 0.75--30 MeV energy band, with photon fluxes in units
of $10^{-4}$ ph cm$^{-2}$ s$^{-1}$ and energy fluxes in units of $10^{-10}$ erg cm$^{-2}$ s$^{-1}$.
}
\end{table*}

\section{Spectral Modelling}
\label{sec:modelling}

To compare the data to expected spectra for positron in-flight annihilation, we modelled the \gray\
emission from a population of energetic positrons interacting with electrons
in the ISM following the prescription in \cite{beacom2006}, \cite{sizun2006} and \cite{prantzos2011}.
Specifically, energy losses through IC scattering, synchrotron emission, Bremsstrahlung
emission \cite[using the prescription of][]{ginzburg1979}, 
Coulomb scattering and ionisation were considered.
While these prescriptions are limited to the case of stationary injection of mono-energetic positrons 
into the ISM, we extend the formalism to arbitrary injection spectra.
The \gray\ emission mechanisms that are included in our modelling are positron in-flight annihilation,
IC scattering, and (external) Bremsstrahlung.
For the case of dark matter annihilation, we also included internal Bremsstrahlung emission following
the prescription of \cite{boehm2006}.
Explicit parameters of our model are the kinetic energy spectrum of the injected positrons, the
hydrogen number density $n_{\rm H}$ of the ISM in which the positrons slow down, and its ionisation 
fraction $Y_{\rm e}$, with $Y_{\rm e}$ = 0 corresponding to the neutral and $Y_{\rm e}$ = 1 to the
fully ionised ISM.
Following \cite{beacom2006}, we normalised the resulting \gray\ intensities in units of 
equivalent 511~keV line flux $\Phi_{511}$ using the observed positronium fraction of 
$f_{\rm Ps}$ = 0.967 \citep{jean2006}, so that when the models are fitted to the COMPTEL data the 
model normalisation is given in units of $\Phi_{511}$.
From $\Phi_{511}$, the positron injection rate $R$ can be derived using
\begin{equation}
R=\frac{4 \pi d^2}{2(1-3/4 f_{\rm Ps})} \frac{\Phi_{511}}{P(E_{\rm inj},E_{\rm therm})}
\end{equation}
where $d$ is the distance to the positron source and $P(E_{\rm inj},E_{\rm therm})$ is the 
integrated survival probability of positrons as they lose energy from the injection energy
$E_{\rm inj}$ to the thermalisation energy $E_{\rm therm}$ \citep{beacom2006}.
Note that $P(E_{\rm inj},E_{\rm therm})$ depends also on $Y_{\rm e}$ and $n_{\rm H}$,
and for $E_{\rm inj} = 1-10$ MeV is in the range within $0.95-0.99$.
For a source at the distance of the Galactic centre \cite[8.2 kpc;][]{gravity2019} this results in
$R \approx 1.5 \times 10^{46} \, \Phi_{511}$,
with $R$ in units of \psec\ and $\Phi_{511}$ in units of \pflux.

We furthermore assumed a positron fraction $N_+/(N_+ + N_-)$ which is used in the computation
of the IC and (external) Bremsstrahlung emissions to account for emission from
both electrons and positrons.
For a pure pair plasma, the positron fraction is $N_+/(N_+ + N_-)$ = 0.5.

We then pre-computed the expected \gray\ intensities for different model parameters and stored
the spectral vectors in so-called `table models'.
The spectral vectors of these models are interpolated when they are used in the model fitting,
which enables determination of, for example, positron injection energies, cut-offs and spectral indices
in a maximum likelihood fit of COMPTEL data.
These models were then used to model the spectrum of either the Bulge or the Core component 
and adjusted jointly over the full $0.75-30$~MeV energy band with an analytical spectral model for 
the other component.
Specifically, a power law spectrum was used for the Core component when table models were 
fitted to the Bulge component, while a log-parabola spectrum was used for the Bulge component when
table models were fitted to the Core component.
In this process, model parameters were fitted and corresponding confidence intervals
assessed, determining in particular positron injection energies and equivalent 511~keV 
line fluxes.

The model spectra shown for the Bulge component in Figs.~\ref{fig:expected-spectra}
were derived for a hydrogen density of $n_{\rm H}=1$ \pcmc\ and an ionisation fraction of 
$Y_{\rm e}$ = 0.1 which corresponds to the value determined from 511~keV line observations 
\citep{prantzos2011}.
The value of $n_{\rm H}$ has little relevance, since it cancels out in the normalisation to the 511~keV 
line flux.

This is no longer true for the model spectrum shown for the Core component in Fig.~\ref{fig:core}, 
since the IC emission, which is dominant in this case, does not depend on the hydrogen 
number density.
We therefore fixed the hydrogen number density to a fiducial value of $n_{\rm H}=1000$ \pcmc\ in 
Fig.~\ref{fig:core} \citep{ferriere2007}.
The ISM is very inhomogeneous in the Galactic centre region, and the actual value to adopt depends
on the medium in which positrons annihilate.
Reducing the value of $n_{\rm H}$ reduces the relative intensity of the positron in-flight annihilation
and Bremsstrahlung emissions, while modifying little the total spectrum that is dominated in any
case by IC emission.
More crucial is the radiation density of the photon field that serves as targets for the electrons and
positrons.
We adopted for Fig.~\ref{fig:core} the R3 radiation field of \cite{hinton2007}, and used again
an ionisation fraction of $Y_{\rm e}$ = 0.1.
The pair plasma was injected into the ISM following a power-law distribution with a minimum kinetic 
energy of 511~keV.
We fixed the 511~keV line flux from thermalised positrons to the observed value of 10$^{-4}$ \pflux\ 
for the central component \citep{skinner2014,siegert2016} and adjusted the maximum kinetic energy
and the power-law slope by jointly fitting the COMPTEL and Fermi/LAT data.
Note that leaving the 511~keV line flux free gave fitted 511~keV line flux values close to the value
observed by INTEGRAL/SPI.

\end{appendix}
\end{document}